\documentclass{PoS}

\title{The Swift UVOT serendipitous source catalogue }

\ShortTitle{The Swift UVOT source catalogue }

\author{\speaker{M. Page}, V. Yershov, A. Breeveld, N.P.M. Kuin, R.P. Mignani, P.J. Smith, \ \ \ \ \ \ \ \ \ \ \ \ \ \ \ \ \ \ \ J. I. Rawlings, S. R. Oates%
\\
       Mullard Space Science Laboratory, UCL, Holmbury St. Mary, Dorking, RH5 6NT, UK\\
       E-mail: \email{m.page@ucl.ac.uk}}

\author{M. Siegel\\
        Department of Astronomy \& Astrophysics, Penn State University, 525 Devey Laboratory, University Park, PA 16802, USA
}

\author{P.W.A. Roming\\
        Southwest Research Institute, 6220 Culebra Road, San Antonio, TX 78238 USA
}

\abstract{We present the first {\em Swift} Ultraviolet/Optical Telescope
  Serendipitous Source Catalogue (UVOTSSC). The catalogue was compiled
  from 23\,059 {\em Swift} datasets taken within the first five years
  of observations with the {\em Swift} UVOT. A purpose-built
  processing pipeline, based around the standard {\em Swift}
  processing tools, was employed. The catalogue contains positions,
  photometry in three UV and three optical bands, morphological
  information and data quality flags. In total, the catalogue contains
  6\,200\,016 unique sources of which more than 2 million have multiple
  observations in the catalogue.}

\FullConference{Swift: 10 Years of Discovery\\
		 2-5 December 2014\\
		 La Sapienza University, Rome, Italy}

\begin{document}

\section{Introduction}

The {\em Swift} Ultraviolet/Optical Telescope UVOT \cite{roming05} is
a 30 cm modified Ritchey-Chretien telescope mounted on the instrument
platform of the NASA {\em Swift} Gamma-Ray Burst space observatory
\cite{gehrels04}. It has many characteristics in common with the {\em XMM-Newton} Optical Monitor 
(XMM-OM) \cite{mason01}, on which its design was based.  A filter wheel provides a selection of lenticular
filters for imaging and two grisms for low-resolution spectroscopy.  A
microchannel-plate intensified CCD (MIC) \cite{fordham89} is used as
the detector. A key characteristic of this detector is an extremely
low dark current. The UVOT is therefore usually background-limited by
the zodiacal light \cite{poole08}. The UVOT has a 17~arcmin $\times$
17~arcmin field of view, so its images routinely contain many more
sources than the target of the {\em Swift} observation. The point
spread function has a FWHM of around 2.5 arcsec in the UV
\cite{breeveld10}.  The combination of the large field of view,
sensitivity in the UV, small PSF and low background mean that the UVOT
is a capable survey instrument.

\begin{figure*}
\begin{center}
\includegraphics[width=\columnwidth]{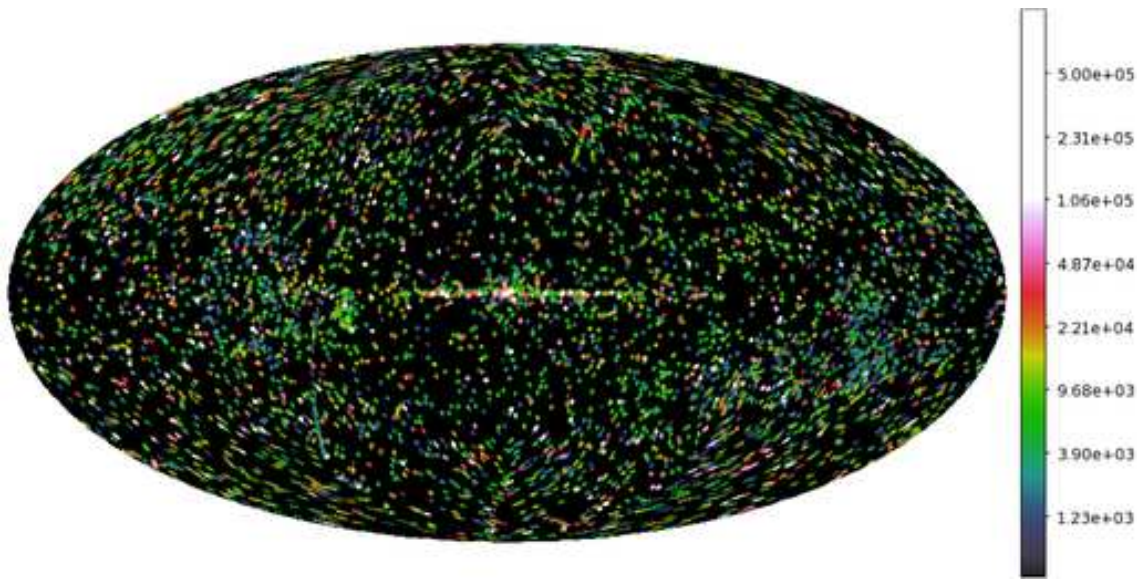}
\end{center}
 \caption{Swift pointings from 2005-2010, colour coded by the total UVOT 
exposure time in seconds} 
 \label{fig:uvotsky}
 \end{figure*}

Swift's observing programme is diverse, from chasing gamma-ray bursts
and other transients \cite{gehrels14}, to pointed observations of all kinds of
astronomical sources. During this observing programme the UVOT has
accumulated a large body of images, sampling a wide range of Galactic
and extragalactic sky; see Fig.~\ref{fig:uvotsky}. 
Since early in the mission, a `filter of the
day' has been chosen for UVOT exposures where there is not a strong
scientific constraint on the UVOT filter. In order to maximise the
serendipitous science return from UVOT, the filter of the day is continuously
cycled through the u and UV filters.


The UVOT Serendipitous Source Catalogue (UVOTSSC) is the first catalogue of
sources detected by UVOT through its six lenticular filters, imaging in
V, B, U, UVW1, UVM2 and UVW2 bands. It is the result of a 5 year long
project led by MSSL, on behalf of the UVOT team, to produce a uniform
product in terms of astrometry, photometry and morphological
information from the first 5 years of UVOT observations. It provides
an easy route for astronomers to obtain UVOT-derived source parameters
without having to reduce or analyse UVOT data, and a large statistical
dataset for the systematic investigation of the UV sky.

\section{Data processing and construction of the catalogue}

\begin{figure*}[tb]
\begin{center}
\includegraphics[width=\columnwidth]{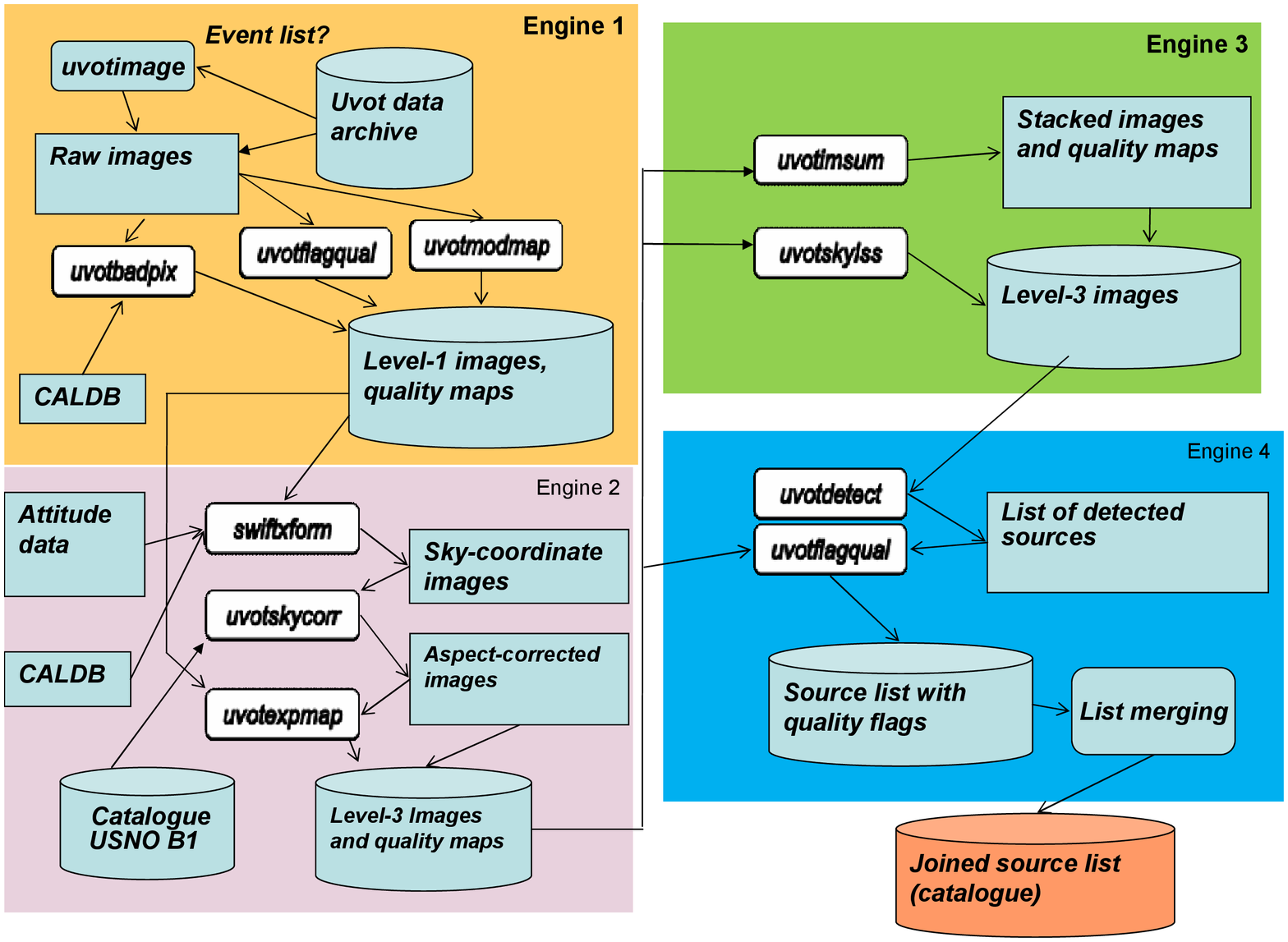}
\end{center}
 \caption{Data processing engines for the UVOT source catalogue} 
 \label{fig:enginesu}
 \end{figure*}

To construct the catalogue, the UVOT data were processed through a
purpose-built pipeline, based on the {\em Swift} UVOT {\sc ftools}
available in {\sc
  heasoft}\footnote{http://heasarc.gsfc.nasa.gov/docs/software/ftools}.
The pipeline is constructed as a sequence of processing engines, which
advance the data through each intermediate stage of catalogue
construction.  The UVOT processing scheme is shown in
Fig.\ref{fig:enginesu}.  The data are processed by {\em Swift}
observation dataset, each of which has a unique 11-digit identifier,
known as OBSID. Each OBSID is processed separately to an
OBSID-specific source list. In the final stage of the catalogue
construction, the individual source lists are brought together.

The vast majority of UVOT observations between 2005 and 2010
inclusive were used as the input data for the catalogue. Very short
exposures 
and
exposures taken when the {\em Swift} star-trackers were not locked
were not used for the catalogue.

\begin{figure*}
\centering 
\includegraphics[scale=0.25]{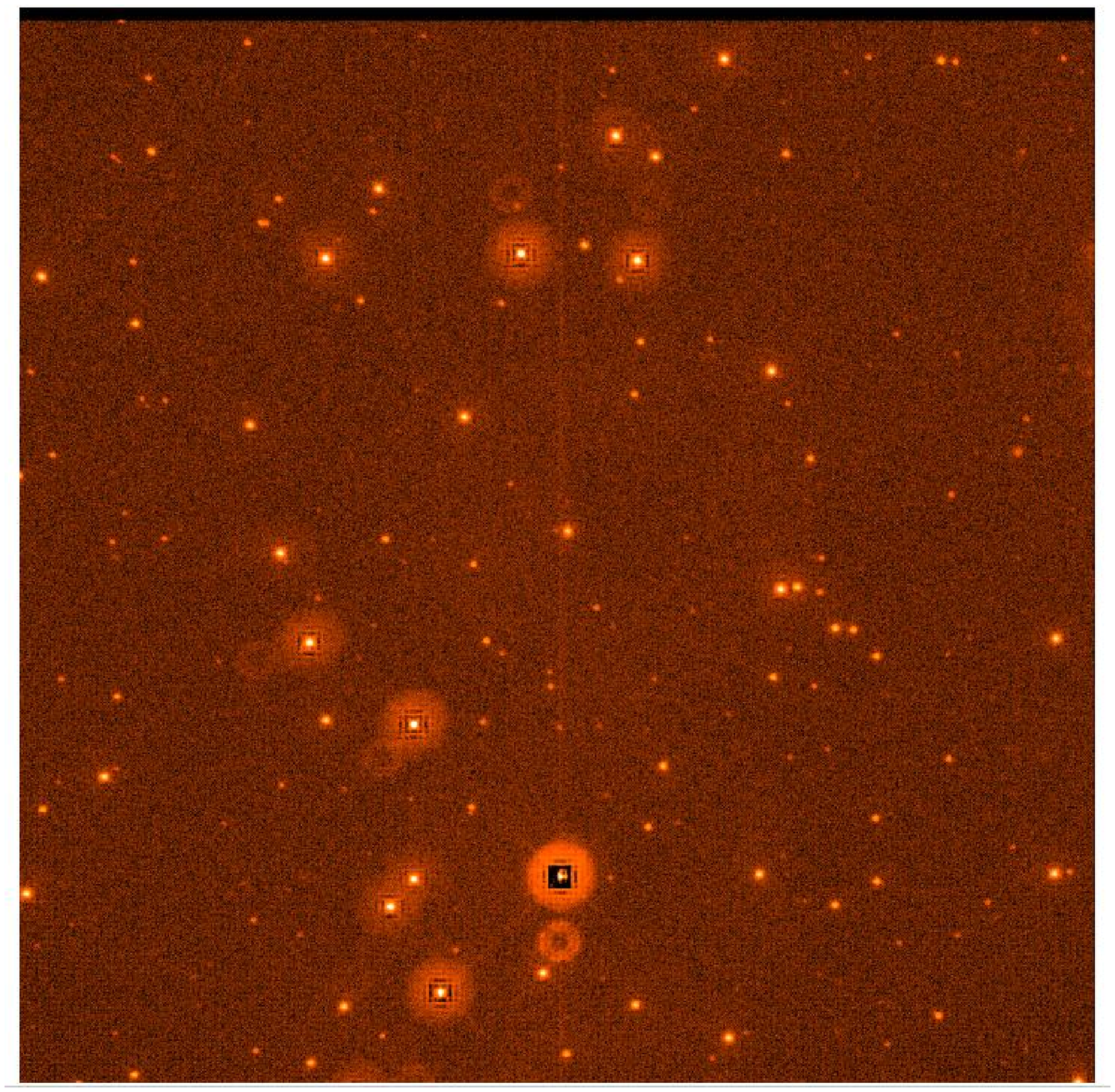}
\includegraphics[scale=0.25]{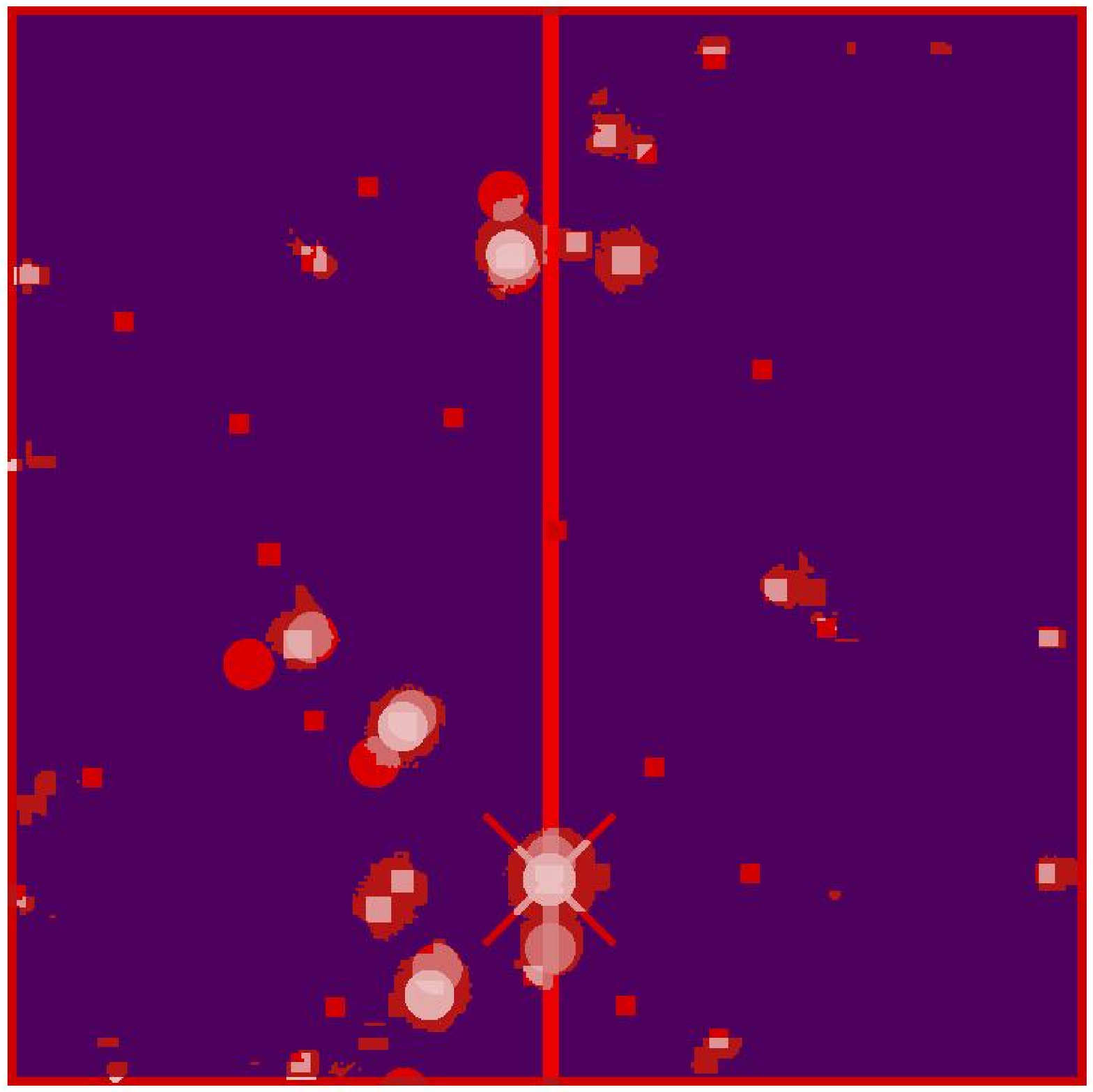}
\includegraphics[scale=0.25]{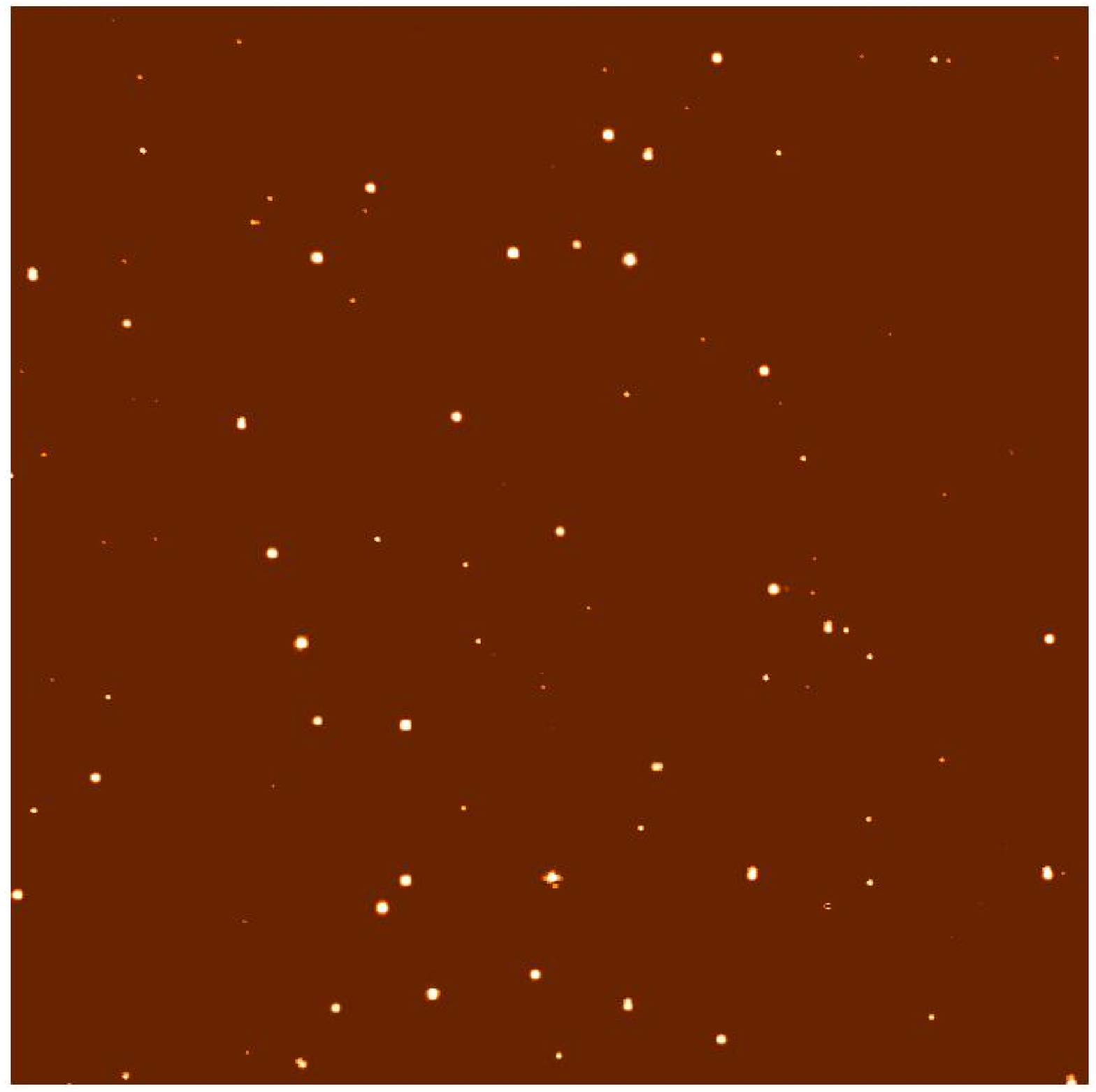}
\caption{Example products of the task {\sc uvotflagqual}: the 
raw image (left panel), the corresponding quality map
(middle panel), and the source map (right panel). The different colours in the quality map correspond to different combinations of quality flags. The source map contains only the pixels corresponding to bright source regions, the rest replaced with the average background of the original image; it is useful if aspect correction of the original image has failed.}
\label{fig:qmap}
\end{figure*}

The first engine creates raw images from data that were taken in
event mode, localises bad pixels (task {\sc uvotbadpix}) and removes
the modulo-8 pattern caused by the on-board centroiding algorithm
(task {\sc uvotmodmap}). The remainder of the first engine deals with
image artefacts (readout streaks, scattered light features, etc.),
creating a map of these artefacts for use later in the processing.
The task {\sc uvotflagqual} identifies possible image artefacts
and sets quality flags in the pixels of the quality map which
accompanies the main image through the further processing steps and which
is used at the final stage of processing for passing quality flags to
those sources whose coordinates coincide with the flagged pixels in the 
quality map.  An example of the quality map containing
a readout streak, diffraction spikes and some other image artefacts is
shown in Fig.\ref{fig:qmap}.  

%

The second engine rotates the images with the purpose of aligning them
along the celestial coordinate axes (task {\sc swiftxform}). Then the 
second engine corrects for the possible shifts of the image
coordinates with respect to the sky coordinates by using reference
stars from the USNO-B1 catalogue (task {\sc uvotskycorr}).  At the end of
this processing stage, the task {\sc uvotexpmap} generates exposure maps
corresponding to the sky-rotated and aspect-corrected images.

The third engine stacks, for each OBSID, the different exposures for each 
filter, as well as the corresponding exposure and quality maps
(task {\sc uvotimsum}) and generates a large-scale sensitivity map needed
for accurate photometry (task {\sc uvotskylss}).

The fourth engine calculates the background maps for each stacked
image and detects sources which have count rates that exceed a
threshold above the background (task {\sc uvotdetect}; 
see Section~\ref{sec:detection}).  The task {\sc
  uvotdetect} used for the catalogue processing was modified with
respect to the standard task in the {\sc heasoft} package in order to deal
with most of the problematic images that could be found in the UVOT
archive, without manual intervention on the parameters of the
task. The final stage of the processing involves the task {\sc
  uvotflagqual} that extracts the image quality flags from the quality
maps produced by the first engine. These flags are introduced into the
source lists for each observation. Finally, when all the UVOT data have 
been processed through the four engines, the source lists are
 concatenated to form the source catalogue and
cross-correlated to identify sources which have been observed in more
than one observation.

\section{Source detection and measurement process}
\label{sec:detection}

For each Swift observation dataset (identified by a unique OBSID
number) and for each filter, all images are processed and then stacked
to achieve maximum sensitivity prior to source detection. Thus within each 
OBSID one image per filter is searched for sources.

The source detection and measurement is carried out within a modified
version of the task {\sc uvotdetect}, which is based around {\sc
  sextractor} \cite{bertin96}. Within {\sc uvotdetect} a background
map is constructed, either using the standard {\sc sextractor}
algorithm which sigma-clips the image until convergence, or for
low-count images using a {\sc uvotdetect}-specific 
algorithm based around Poisson
statistics. Sources are then detected by blurring with the standard
{\sc sextractor} pyramidal function before searching for groups of 8
or more connected pixels which are brighter than the background by
more than 1.5 sigma.

Source countrates and magnitudes are calculated as corrected isophotal
magnitudes in {\sc sextractor} terminology \cite{bertin96}. Only
detections with a signal to noise ratio $> 3$ are retained after the
source detection passes. Quality flags are then propagated from the
quality maps to the source lists. Within each OBSID, the source lists
are then merged to form a single source list per observation, one row
per source, which contains the photometric, morphological and quality
information from all the UVOT filters used in that OBSID. From this list,
any sources which do not have a signal to noise ratio $>5$ in at least
one filter are removed.

\section{Catalogue structure}
\label{sec:catstruct}
      
The UVOT source catalogue is presented in the form of FITS files with
two table extensions, the first table (called SOURCES) containing the
source parameters, and the second table (SUMMARY) containing
information about the observations used for producing the source
catalogue. The catalogue has deliberately been given a similar
structure to the XMM-SUSS catalogue compiled from XMM-OM images
\cite{page12}, given the similarity in the character of the data
contained within the two catalogues.

Since the images from different observations (OBSIDs) generate separate
source lists that are concatenated in the final stage of catalogue production, 
the same source could be detected in several different
observations.  Such sources have multiple entries in the final
catalogue SOURCES table, but each entry has the same unique source 
identification
number, so the number of entries in the final source list table is
larger than the total number of sources in the catalogue.

The first column (IAUNAME) of the SOURCES table gives the IAU source
name in the form {\sf SWIFTUVOT JHHMMSS.S+DDMMSS}, where HHMMSS.S stands for
right ascension coordinates, and DDMMSS for declination coordinates. 
The second column
(N\_SUMMARY) contains a number which links the SOURCES and SUMMARY
tables with each other, that is, for each source it gives the row in
the SUMMARY table corresponding to the observation in which the source
was detected. The third column (OBSID) contains the observation number
within the {\em Swift} archive in which the source was detected.  The
fourth column (NFILT) gives the number of UVOT filters through which
the source was observed. The next four columns contain the equatorial
coordinates and position uncertainties for the source. The next six
columns ({\em filter} SRCDIST), where {\em filter} corresponds to the
UVOT filter names, give the distance to the nearest neighbouring
source detected in the corresponding filters. This is useful to assess
whether the source parameters are likely to have been affected by the
proximity of (and confusion with) other sources. The next column
(NOBSID) reports the number of {\em Swift} observations in which the
source has been detected. The next 30 columns report the photometric
properties of the source in each of the filters through which it has
been observed: signal to noise ratios, magnitudes in the AB and Vega
systems together with the associated uncertainties, flux estimates and
the corresponding uncertainties. The next 24 columns contain
morphological information by filter: major and minor axis sizes,
position angles and whether the source is considered extended. The
final six columns provide quality flags, broken down by filter.

\section{Quality flags}
\label{sec:qualityflags}

\begin{table}
  \centering
    \caption{Source quality flags within the UVOTSSC.}
    \label{tab:qualityflags}
    \begin{tabular}{cll}
      \hline
      Bit & Value & Quality issue\\
      \hline
      0 & 1   & Cosmetic defects (bad pixels) within the source region.\\
      1 & 2   & Source lies on or near a bright read-out streak\\
      2 & 4   & Source lies on or near a smoke ring\\
      3 & 8   & Source lies on or near a diffraction spike\\
      4 & 16  & Source is bright with coI-loss-induced mod-8 noise\\
      5 & 32  & Source lies within a `halo ring' of enhanced background\\
      6 & 64  & Source lies close to a bright object\\
      7 & 128 & Source lies close to a large change in the exposure map \\
      8 & 256 & Point source lies over an extended source\\
      \hline
    \end{tabular}
\end{table}

An important part of the catalogue processing is the recording of any
issues that may affect the quality of the catalogue data so that the
user can make an informed decision as to how reliable those data
are. In the UVOTSSC pipeline this information is encapsulated in the
form of quality flags which are propagated from the first processing
engine to the final catalogue, on a source by source, filter by filter
basis. These quality flags are stored as an eight-bit
integer number, with each bit corresponding to one of the quality flags. 
Any bits which are set
(i.e. non-zero) indicate a potential quality issue. Table
~\ref{tab:qualityflags} lists the quality flags, the values of the
bits that they represent and a brief description of each flag.

\begin{figure*}
\begin{center}
\includegraphics[width=\columnwidth]{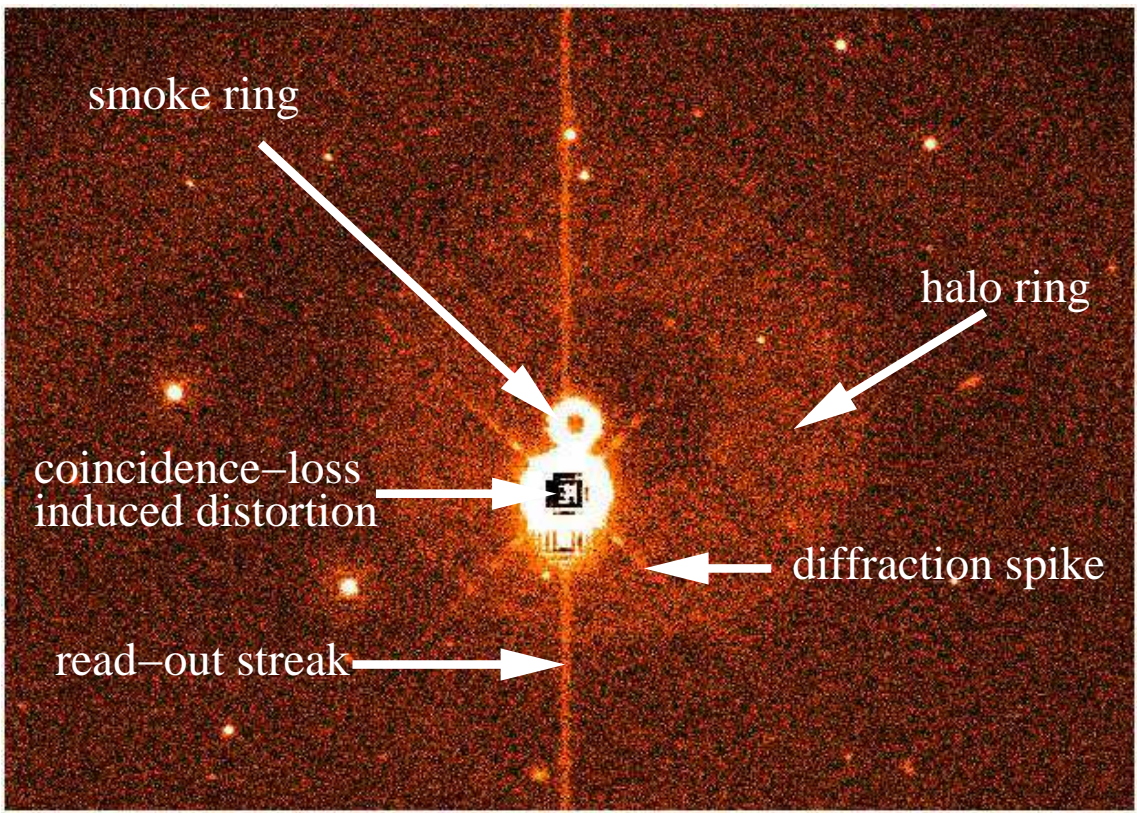}
\end{center}
\caption{An image containing an example of a read-out streak,
  coincidence-loss induced distortion, a smoke ring and a halo
  ring, around a bright source which shows diffraction spikes. The 
different features are labelled.}
 \label{fig:ringsnstreaks}
 \end{figure*}

Some of the quality flags are generic, while some are particular to the
UVOT and its sister instrument, the XMM-OM, and will be described
briefly here. The CCD that forms the last stage of the UVOT MIC
detector is a frame-transfer device, and photons which arrive while
the frame is being transferred to the frame store will be displaced in
the vertical direction \cite{page13}. Bright stars thus give rise to vertical
streaks of displaced photons which can be seen in UVOT images. These
streaks are referred to as read-out streaks. Occasionally, {\sc
  uvotdetect} detects spurious sources associated with the read-out
streaks, and very faint sources close to the read-out streaks may have
their photometry affected by the presence of the streaks.

The UVOT is a photon counting instrument, and is linear in the
asymptotic limit of faint sources. However, when the count rate from
an object is an appreciable fraction of the frame rate (90.6 frames
s$^{-1}$ in normal, full-frame operation), more than one photon may
arrive within the same frame, and will only be counted as a single
photon, resulting in a non-linearity of the detector. This is called
coincidence loss \cite{fordham00} and is analogous to pile-up in X-ray CCDs. A
correction for coincidence loss is applied in the calculation of the
photometry, but for sources with large coincidence loss the image
becomes distorted with a modulo-8 pattern relating to the
event-centroiding that can not be corrected. In particular, sources
develop a dark ring around a bright core. Morphological information is
compromised for such sources.

Smoke rings and halo rings are associated with bright sources, and are
produced by internal reflections within the detector window. Smoke
rings are compact (30 arcsec diameter), out-of-focus images of the
source displaced radially from the in-focus image, while the optical
path responsible for the halo rings produces much larger (4.5
arcminute diameter) features. Smoke rings may give rise to spurious
sources and may contaminate the photometry of nearby sources. Halo
rings, with their larger scales, produce an enhanced and sometimes
spatially variable background, and so will affect the photometry of
faint sources. Fig.~\ref{fig:ringsnstreaks} shows examples of a
read-out streak, coincidence-loss induced distortion, a halo ring and
a smoke ring around a bright source with diffraction spikes visible.

\section{Properties of the catalogue}

\begin{table}[h]
\caption{UVOTSSC overall statistics}
\label{tbl:catstat}
\begin{center}
\begin{tabular}{ll}
\hline  
Period of observations    &  2005--2010    \\
Total observations     &  23\,059  \\
Total sources     &  6\,200\,016 \\
Repeated observations   & 2\,027\,265  \\
Total entries      & 13\,860\,569  \\
\hline 
\end{tabular}
\end{center}
\end{table}

\begin{figure*}
\begin{center}
\includegraphics[width=11cm,angle=-90]{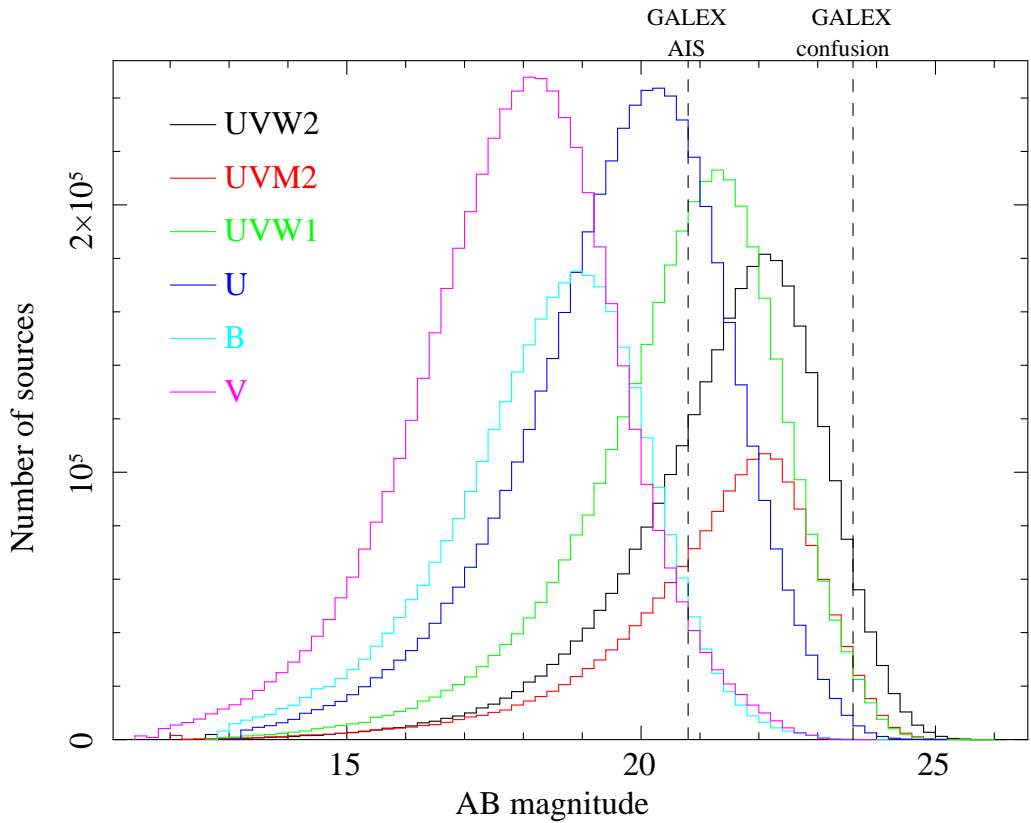}
\end{center}
\caption{Distributions of source magnitudes in the six imaging filters used for the UVOTSSC. The GALEX All-sky Imaging Survey (AIS) depth, and the GALEX near-UV confusion limit are indicated.}
 \label{fig:magnitudes}
 \end{figure*}

Some simple statistics about the catalogue as a whole are summarised
in Table~\ref{tbl:catstat}.  Statistics on the UVOTSSC, broken down by
filter, are provided in Table~\ref{tab:filterstat}, and the
distribution of source magnitudes in the six filters is shown in
Fig.~\ref{fig:magnitudes}. The bright end limit of the catalogue
corresponds to the limit at which coincidence loss can no longer be
corrected \cite{page13}. The catalogue typically reaches fainter
magnitudes in the UV filters than the optical filters, and the
majority of the UV sources are fainter than the {\em GALEX} All-Sky
Imaging Survey limit \cite{morrissey07}. A significant minority of
UVOTSSC sources are fainter than the {\em GALEX} confusion limit
\cite{xu05}. The UVOTSSC is a factor 2 larger than the second release
of the XMM-SUSS \cite{page12}, reaches about 1.4 magnitudes deeper
than the second release of the XMM-SUSS in the UVM2 filter, and 2
magnitudes deeper in UVW2.

\begin{table}
\caption{UVOTSSC statistics by filter}
\label{tab:filterstat}
\begin{center}
\begin{tabular}{lllllll}
\hline  
&V&B&U&UVW1&UVM2&UVW2\\
\hline
Total detections        &5\,087\,552 &3\,329\,392 &4\,931\,791 &3\,832\,449 &1\,799\,025 &3\,110\,521 \\
Mean AB magnitude       &17.81 &18.37 &19.65 &20.62 &21.16 &21.38 \\
Mean magnitude error    &0.08 &0.09 &0.09 &0.10 &0.13 &0.11 \\
Brightest AB magnitude    &11.40 &12.83 &13.18 &12.80 &12.06 &12.63 \\
Faintest AB magnitude    &23.82 &24.18 &25.09 &25.86 &25.70 &26.00 \\
\hline 
\end{tabular}
\end{center}
\end{table}

\subsection{Astrometry}

UVOT astrometry is tied to the USNO-B1 catalogue during the data
processing. The absolute astrometric accuracy of UVOT-derived
positions is better than 0.5 arcsec \cite{breeveld10}, but for
the faintest sources the statistical uncertainty on position is
comparable to the systematic uncertainty, and for large extended sources the uncertainties can be
larger. Within the catalogue, 98 per cent of the sources have
statistical uncertainties on their position of less than 0.5 arcsec.

\subsection{Photometry}
\begin{figure*}[tb]
\begin{center}
\includegraphics[width=\columnwidth]{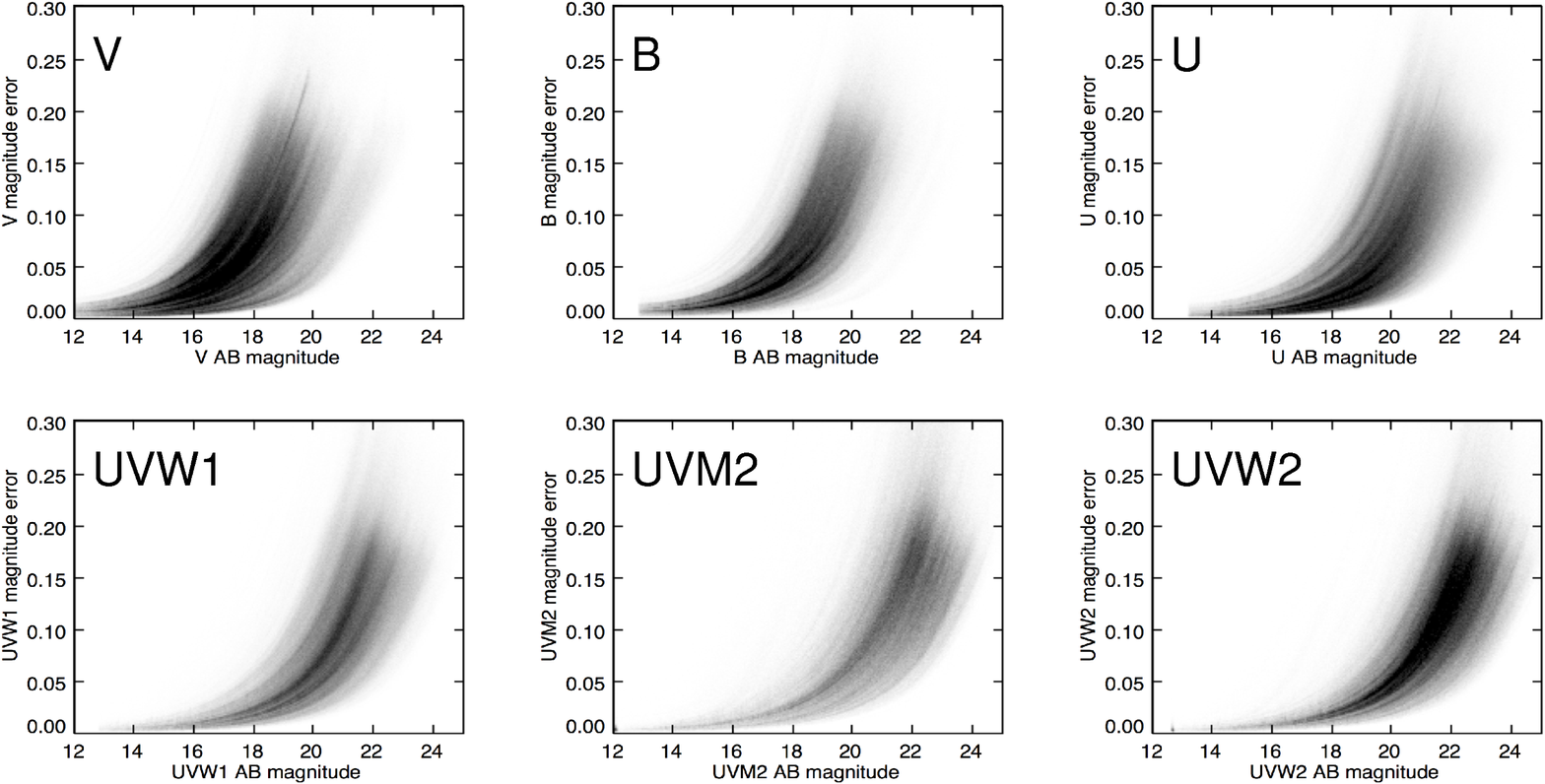}
\end{center}
 \caption{Magnitude errors against AB magnitude for the 6 photometric 
bands used in the catalogue.} 
 \label{fig:magnitude_errors}
 \end{figure*}

The photometric calibration of UVOT is described in \cite{poole08}, with updates 
in \cite{breeveld10} and \cite{breeveld11}. 
Within the catalogue, the photometric accuracy has a strong dependence on source
magnitude, but also depends on exposure time (which has a wide range) and background. 
The distribution of photometric uncertainty is shown
against magnitude for the six photometric bands in
Fig.~\ref{fig:magnitude_errors}.

\section{Accessing the catalogue}

The UVOTSSC can be downloaded from the MSSL {\em Swift} UVOT web
pages\footnote{http://www.ucl.ac.uk/mssl/astro/space\_missions/swift/uvotssc}. It
has also been delivered to the Mikulski Archive for Space Telescopes
(MAST)\footnote{http://archive.stsci.edu} where it will be available
soon. In due course, the UVOTSSC will also be available through the
Virtual Observatory.

\section{Conclusions}

We have constructed a catalogue of objects detected in {\em Swift}
UVOT images. The catalogue was constructed via a purpose-built
processing pipeline based around the standard UVOT {\sc ftools}.  The
UVOTSSC contains 6\,200\,016 unique sources detected with the {\em
  Swift} UVOT within the first 5 years of {\em Swift} operations. For
each source, astrometry, morphology and photometry in up to six UV and
optical bands is provided, together with information about the quality
of the data. Now that the machinery is in place to process the
catalogue, we intend to add the second five years of UVOT observations
for the next release.  The catalogue offers a large statistical sample
to explore the properties of Galactic and extragalactic UV source
populations, and a convenient means for astronomers to obtain
UVOT-derived information about sources of interest without having to
reduce UVOT data.


\begin{thebibliography}{99}

\bibitem{roming05}
Roming P.W.A., et~al., 2005, Space Sci. Rev., 120, 95

\bibitem{gehrels04}
Gehrels N., et~al., 2004, ApJ, 611, 1005

\bibitem{mason01}
Mason K.O., et~al., 2001, A\&A, 365, L36

\bibitem{fordham89}
Fordham J. L. A., Bone D. A., Read P. D., Norton T. J., 
Charles P. A., Carter D., Cannon R. D., Pickles A. J., 
1989, MNRAS, 237, 513

\bibitem{poole08}
Poole T.S., et~al., 2008, MNRAS, 383, 627

\bibitem{breeveld10}
Breeveld A.A., et~al., 2010, MNRAS, 406, 1687 

\bibitem{gehrels14}
Gehrels N., Cannizzo J.K., 2014, arXiv:1502.03064 

\bibitem{bertin96}
Bertin E., Arnouts S., 1996, A\&A Supp. 317, 393

\bibitem{page12}
Page M.J., et~al., 2012, MNRAS, 416, 2792

\bibitem{page13}
Page M.J., et~al., 2013, MNRAS, 436, 1684

\bibitem{fordham00}
Fordham J.L.A., Moorhead C.F. \& Galbraith R.F., 2000, MNRAS, 312, 83

\bibitem{morrissey07}
Morrissey P., et~al., 2007, ApJS, 173, 682.

\bibitem{xu05}
Xu C.K., et~al., 2005, ApJ, 619, L11

\bibitem{breeveld11}
Breeveld A.A., Landsman W., Holland S.T., Roming P., Kuin N.P.M., Page M.J.,
2011, AIP Conference Proceedings, 1358, 373


\end{thebibliography}
\end{document}